\documentclass[12pt]{iopart}

\usepackage{cite}
\usepackage{graphicx}
\usepackage[dvipsnames]{xcolor}
\usepackage{soul} \sethlcolor{CarnationPink}
\begin{document}

\title[Modeling helium compression and enrichment in DIII-D edge plasmas]{Modeling helium compression and enrichment in DIII-D edge plasmas using the SOLPS-ITER code}

\author{R. Masline*, M. Wigram, D. Whyte}

\address{Massachusetts Institute of Technology Plasma Science and Fusion Center, Cambridge, Massachusetts 02139}
\ead{rmasline@psfc.mit.edu}
\vspace{10pt}
\begin{indented}
\item[]\textit{* Supported by Oak Ridge Institute for Science and Education (ORISE) managed for the U.S. Department of Energy Fusion Energy Sciences Postdoctoral Research Program.} \\
\item[]2025
\end{indented}

\begin{abstract}
    Efficient removal of helium ash is a critical requirement for the operation of fusion power plants, as its accumulation can dilute the core fuel and degrade plasma performance \cite{Krasheninnikov_1991}. While past studies suggested that helium exhaust in burning plasmas could be managed effectively through divertor optimization and conventional cryopumping, a detailed understanding of helium behavior in the edge and divertor plasma remains limited, as helium transport through the edge plasma is complex and fundamentally different from other impurity species. With the emergence of more sophisticated numerical modeling tools and renewed focus on D-T burning plasmas, revisiting helium transport in current magnetic confinement devices is necessary for planning and designing fusion pilot plants. This study uses SOLPS-ITER  \cite{schneider} to model a helium-seeded discharge from the DIII-D tokamak \cite{Wade_1998}, analyzing the transport, recycling, and enrichment of helium in the divertor. In addition to characterizing helium dynamics, the results are interpreted in terms of the Tritium Burn Efficiency (TBE) \cite{tritium_memo}, a recently proposed metric linking helium exhaust fraction to tritium fuel utilization in steady-state burning plasmas. By assessing the compatibility of TBE assumptions with detailed edge plasma simulations, this work provides insight into the practical viability of TBE as a reactor design and performance metric.
\end{abstract}

%
%
\submitto{\NF}
%
%
%

\section{Introduction and Motivation}
Fusion power plants will need to contend with constant production of helium ash, born as alpha particles that are produced as a byproduct of fusion reactions. This helium ash originates in the core plasma and must cross the separatrix, pass through the divertor plasma, neutralize on the wall or divertor target, and enter the pumping plenum to be removed as atoms. Otherwise, a build-up of ash in the core and exhaust stream can contaminate the fusion fuel and degrade confinement, making efficient removal of helium a key priority for power plant design and operation. The focus of this work is to study the behavior of the helium species after it crosses the separatrix, where it is subject to the complex dynamics of the edge and divertor plasma \cite{Krasheninnikov_1991}.

The helium species itself is unique in its atomic properties, giving it many interesting features that are distinct from the main ion hydrogenic species or other impurity species. Compared to deuterium, helium has an extremely high ionization energy, does not form molecules, and has a very small cross section, experiencing very little interatomic or intermolecular charge exchange interaction due to the close proximity of the filled valence electron shell to the nucleus \cite{Kubo_1999}, but can undergo elastic collisions with the background species that can significantly impact helium transport due to scattering since helium has a similar mass (unlike other impurity species). Because of these reasons, helium recycles differently, entrains in the plasma flow differently, and behaves differently with respect to the conductance of the atomic helium species in the exhaust plenum than any other species found in the edge plasma. It is not generally considered a radiating impurity species, but can dilute the core plasma and throttle or even collapse the overall reactivity of the plasma fuel if it is allowed to accumulate in the system.

Despite the unique characteristics and importance of removal and management of helium in the edge plasma, the concern of helium in the exhaust stream has largely been ignored in recent years. There has been a a general sense of confidence in meeting the requirements This reduced focus appears to stem from a general confidence established by the extensive simulations and analyses conducted by Kukushkin and others for the ITER tokamak, where meeting the requirements for efficient helium ash removal was a central consideration in the divertor design \cite{Kukushkin_2002,ITERPhysicsExpertGrouponDivertor1999, Gilleland_1989}. In addition, multiple experimental studies and corresponding simulations in the 1990s explored helium compression and enrichment in JET, ASDEX, JT-60, DIII-D, and other machines \cite{hillis1990,hillis1999,Groth2002,Wade_1998,BOSCH1997,Bosch_1997,coster2001,GOETZ1999,Ishida_1999}. Initial estimates based on simulations of ITER showed that with optimization of target shape and divertor structures and including realistic models of elastic collisions between the main ion species and helium, plasmas in ITER would achieve the necessary pumping requirements to avoid helium accumulation in operational, partially detached burning plasmas \cite{Kukushkin_2002}. Experimental results from this time period supported this, demonstrating exhaust enrichment values of 0.2 to 1, which was considered manageable within the constraints of the fusion-relevant plasmas anticipated in the ITER device at that time \cite{hillis1999,Groth2002,wade1995,Wade_1998,Bosch_1997}. However, numerical modeling of these experiments was extremely limited \cite{hillis1990,hillis1999,Groth2002,Wade_1998,BOSCH1997,Bosch_1997,coster2001,Kukushkin_2002} or never done \cite{GOETZ1999,Ishida_1999}, making the predictive capabilities of the existing modeling tools for helium in the edge plasma largely unknown.

With the fusion community now looking towards burning D-T plasmas not just for science progress, but also to produce commercial fusion, there is renewed interest in studying helium exhaust using existing magnetic confinement devices, as these plasmas will be capable of generating considerable amounts of helium ash. New technologies, both in terms of new machine designs enabled by advancements in hardware and considerable maturity in the development of numerical modeling codes, raise questions about the applicability of the existing basis of the understanding of helium transport and enable interpretive studies with higher fidelity than the previous models. An extensive study of helium dynamics was performed for ASDEX with both experiments and numerical modeling \cite{zito_thesis, Zito_2023, Zito_2025}, illuminating many of the physical properties determining the behavior of helium in the plasma exhaust stream and the limitations in the numerical models used to characterize the helium species. Much of the analysis and conclusions presented in the work here corroborate the findings of \cite{Zito_2025}, and echo the call presented by these authors that further study and understanding of helium exhaust in different operating regimes, machines, and divertor configurations is absolutely necessary to lay the foundations for successful operation of future burning plasma experiments and eventual power plants. 

\begin{figure}
    \centering
    \includegraphics[width=\linewidth]{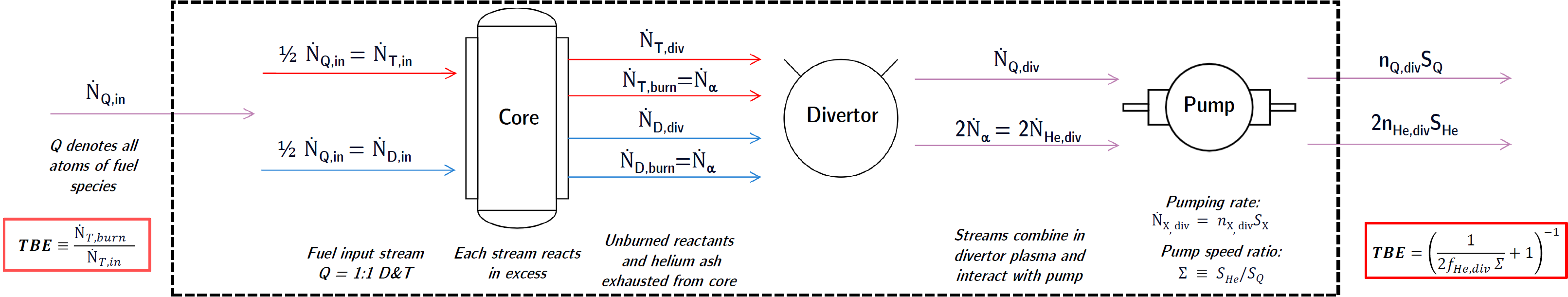}
    \caption{Tritium Burn Efficiency (TBE) \cite{tritium_memo} connects steady state fueling, pumping rate, and helium ash, treating the reactor plasma and divertor system through particle balance, as shown.}
    \label{fig:TBE}
\end{figure}

These advanced numerical models that enable better interpretation of helium transport in the edge plasma that serve as validation exercises for real experiments are essential to understanding and interpreting the predictive capabilities of these codes, which can be used for future reactor designs, projections of possible plasma operating scenarios, or validation of analytical scalings and formulations. One such application is relevant for characterizing the fuel cycle - in addition to the renewed interest in understanding the physical dynamics of helium exhaust, recently, it was proposed that the helium exhaust and the usage of tritium in steady-state equilibrated fusion power plants could be characterized by a generic, dimensionless figure of merit: the “Tritium Burn Efficiency”, or “TBE” \cite{tritium_memo} which connects the performance of the power plant with the permitted helium gas fraction in the divertor. In this formulation, the TBE parameter is effectively a proxy for fusion performance: as illustrated in Figure \ref{fig:TBE}, the concentration of helium in the exhaust stream can act as a surrogate for the total number of fusion reactions occurring in the system, since at steady-state, each fusion reaction will yield an alpha particle at the same rate as helium ash atoms are removed. By the assumption of steady-state conditions of the output stream, rather than the input, this parameter provides a means of interpreting the actual consumption rate of tritium without requiring explicit specification of fueling rates or residence times when evaluating fuel efficiency, unlike previously defined metrics for quantifying projected fuel usage \cite{Abdou_2021}. These quantities (fueling rates and residence times) can only describe a single pass of tritium through the plasma and do not fully capture the complex dynamics of divertor recycling and recirculation that are inherently accounted for in the steady-state assumptions of the TBE model. Since the TBE model draws a very powerful connection between measurable divertor quantities and the overall performance of any steady-state fusion power plant, agnostic of magnetic configuration, it is an appealing metric for a power plant designer - however, the model is simple in that it says nothing at all about the actual boundary plasma conditions (such as flow, ionization, recycling, or pump technology) of what actually sets compression and pumping of helium, which are the determining factors for the TBE metric. Edge plasma simulations, which incorporate complex recycling physics, steady-state assumptions, and realistic considerations for pumping, can provide a thorough assessment of the viability and optimization of this parameter, and namely whether realistic conditions for the boundary plasma and pumping throughout can be achieved with reasonable upstream parameters.

In this work, a discharge from the DIII-D tokamak with an extrinsic injection of helium will be modeled using the SOLPS-ITER code \cite{schneider} and analyzed. The goal of this study is twofold: first, the behavior, transport, and recycling dynamics of the helium species will be described, and second, these results will be discussed and analyzed in terms of the TBE parameter to assess the viability of this parameter as a design metric for experimental scenarios and future power plants. Section \ref{experiment} describes the experiment and corresponding simulation setup for the DIII-D discharge modeled in this work, Section \ref{comparison} discusses the matches between the reported experimental measurements and the simulation, Section \ref{results} shows the results of the analysis of helium transport, and Section \ref{tbe} discusses the results of this modeling work in terms of its applications to TBE.

\section{Experiment and simulation setup}
\label{experiment}

\begin{figure}
    \centering
    \includegraphics[width=0.5\linewidth]{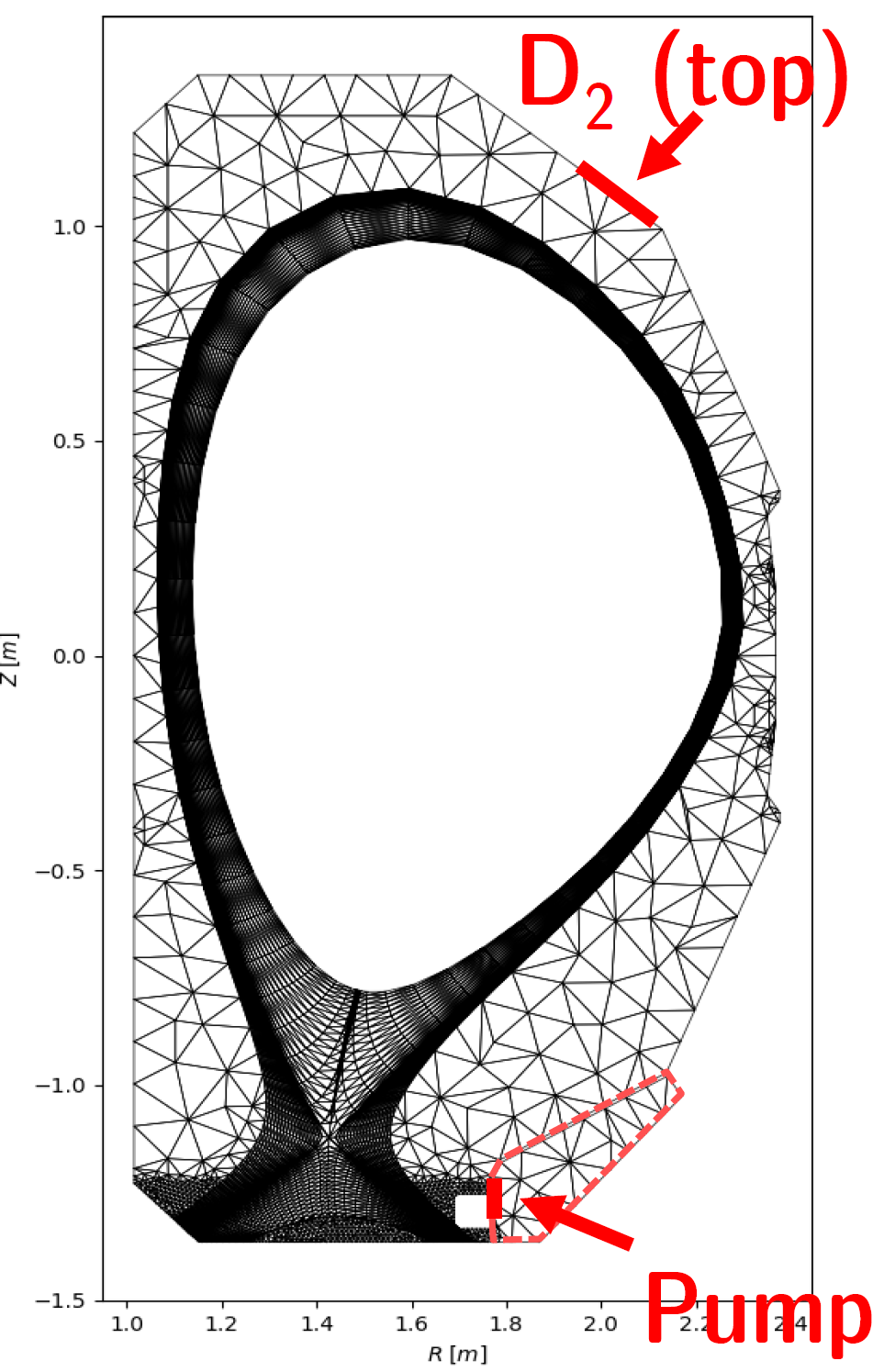}
    \caption{Computational grid used in the SOLPS-ITER simulation, showing the B2.5 plasma mesh (curvilinear cells) and the extended EIRENE domain (triangular cells) used to model a realistic divertor pump volume. The extended EIRENE region allows for more accurate characterization of neutral particle transport and helium exhaust behavior, particularly within the divertor plenum and pump slot geometry.}
    \label{fig:diiid-grid}
\end{figure}

Since the availability of experimental data for tokamak discharges using helium impurity is limited, the parameters matched in this study come from the profiles and quantities reported in existing work from DIII-D in the mid-1990s \cite{wade1995,Wade_1998}. The motivation for these experiments, dubbed ``puff and pump" studies, was to better understand the impact of scrape-off layer flows on impurity compression and enrichment. To do this, deuterium was injected via gas puff in either an upper valve on the low field side (marked in Figure \ref{fig:diiid-grid}) or in the divertor to compare the behavior of injected helium, neon, and argon impurities between regimes with and without strong induced flows in the SOL. The experiments revealed significant increases in impurity compression and enrichment within the divertor when SOL flow was induced, and that the degree of enrichment varied depending on the specific impurity species, indicating a dependency on atomic characteristics. 

\begin{figure}
    \centering
    \includegraphics[width=0.5\linewidth]{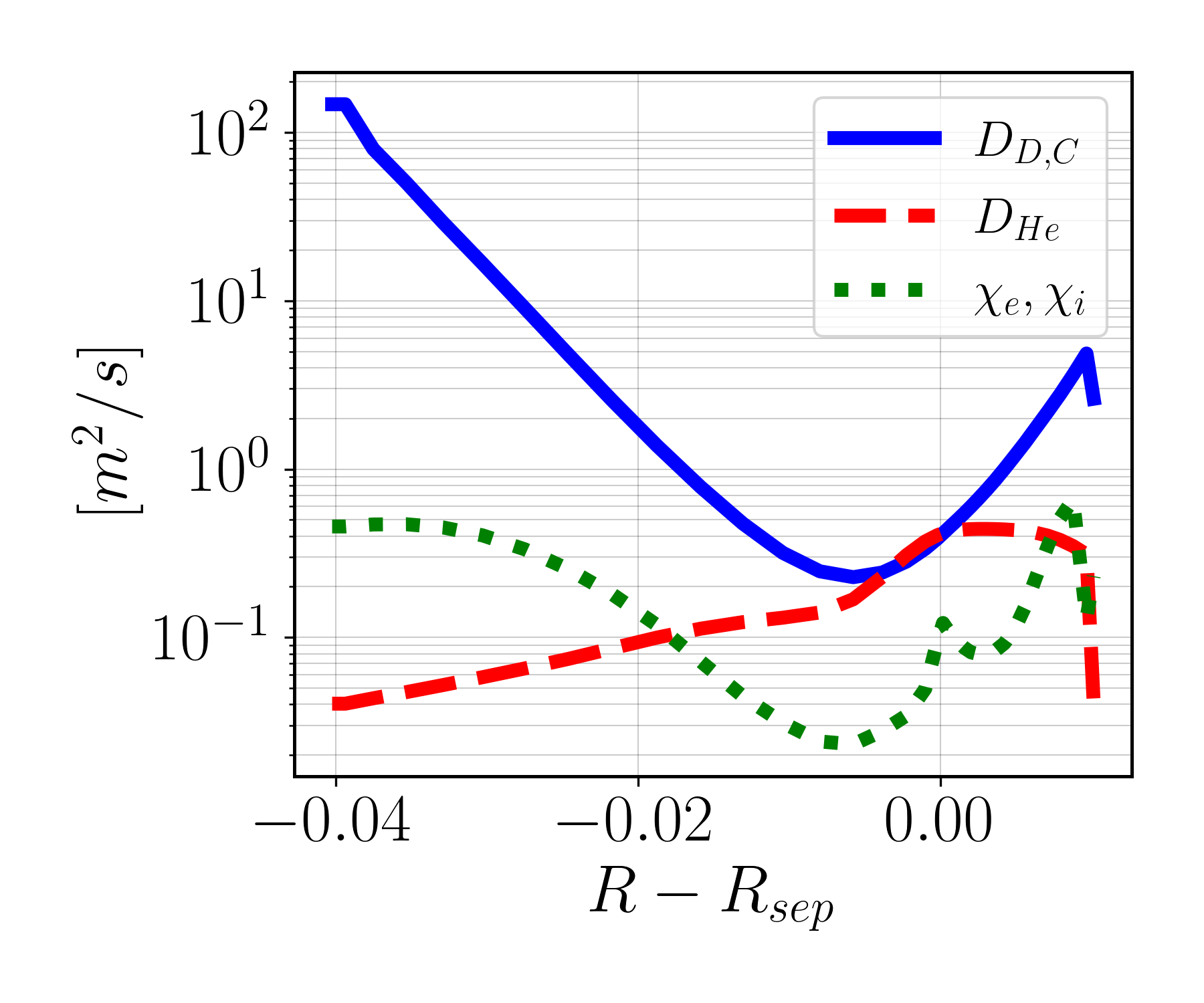}
    \caption{Radially varying transport profiles for the diffusion coefficient D for deuterium and carbon (blue), diffusion coefficient D for helium (red dash), and heat diffusivity $\chi$ for all ion species and electrons (green tick).}
    \label{fig:coefficients}
\end{figure}

In an effort to understand the transport of helium in the edge plasma, the input parameters used to constrain the simulation were carefully tuned to match the limited quantity of experimental data with the simulation results. Specifically, the simulations were tuned to match the quantities presented in Table II of \cite{Wade_1998} in the divertor, and the figures presented in that work. Raw experimental data signals for these discharges are available from the DIII-D MDSPlus database, but processed signals are not available; instead, experimental data profiles were reconstructed using ``best fit" parametrization from data points extracted from the figures displaying the processed data shown in the paper and corresponding data files were prepared that were used as inputs for different tuning routines and modules developed for the SOLPS-ITER code framework to characterize the radial transport coefficients (shown in Figure \ref{fig:coefficients}) necessary to reproduce these profiles. In particular, this technique was applied using the SOLPSXport tool with an additional custom module developed for this work to separately tune and match the helium density gradients and applied iteratively in conjunction with modifications of the leakage boundary condition setting \cite{CANIK2011,DEKEYSER2017}. This technique was applied to the electron density, electron temperature, and core helium profiles, for which transport profiles were individually tuned and applied separately to both the $He^{1+}$ and $He^{2+}$ species. Measurements for the SOL profiles of helium were extrapolated from the very last radial CER measurement with an assumed hyperbolic tangent structure across the SOL that matched the shape of the main ion species, yielding a density profile consistent with the assumed profile in the MIST modeling impurity transport modeling shown in the paper. It is noted that the far SOL quantities of helium are unknown, but this set of upstream parameters corresponding to the ``synthetic" profile for the helium ions yielded reasonable conditions downstream that were then separately further tuned to result in very good agreement with the reported experimental measurements of helium pressure in the exhaust duct. The fueling rate and input power are set constant based on the values reported in the original paper. 

Since the initial work placed emphasis on the consistency in neutral pressure and particle balance in the experiment, neutral pressure is one of the key constraints in this simulation and it was paramount that the simulation represent the most accurate-possible characterization of the pumping plenum for DIII-D. The simulation was constructed using a vessel contour corresponding to DIII-D from the 1990s. Typically, in simulations of DIII-D, the entrance to the pumping plenum is the set as the effective pumping surface and the entire volume of the plenum is not included. Therefore, the vessel contour was modified to an include an additional structure that was added to the computational domain based on schematics \cite{cryopump} and the reported pumping plenum volume of 1.3 $m^{3}$ stated in the initial work \cite{Wade_1998}. This was achieved by constructing a new EIRENE mesh with the approximate specifications shown in the schematics, comparing (assuming 3D toroidal symmetry) the total volume of the new domain with the plenum to the old one without the plenum, and then iterating on the precise dimensions of the walls of the plenum volume in the vessel contour to achieve the specific pump plenum volume reported in the paper, shown in Figure \ref{fig:diiid-grid}.

\section{Comparison to Experiment}
\label{comparison}
The parameters matched between the experiment and simulation are detailed in Table \ref{tab:table_compare}. For this simulation, the exhaust helium fraction $f_{He,exh}$ is defined as the ratio of the number of helium neutral atoms in a volume-averaged region of the EIRENE computational domain in the plenum volume relative to twice the deuterium molecular density in the same region: $n_{He}/2n_{D2}$. The core helium fraction $f_{He,core}$ is defined as the ratio of the sum of $He^{1+}$ and $He^{2+}$ relative to the ionized deuterium density in the computational cell just outside the separatrix at the midplane: $n_{He^{[1-2]+}}/n_{D^{+}}$. For the calculations of compression, denoted as $C_{He,exh}$, some reinterpretation of the reported and calculated values is necessary. It is noted that the published compression ratios for all 10 discharges involving He, Ne, and Ar in detailed in \cite{Wade_1998} are systematically incorrect; specifically, they are $\sim$30\% higher than the values obtained by directly calculating the compression from the published upstream and downstream density values listed in the same tables. The value for helium compression calculated from the published density values is listed in the ``Corrected" row of Table \ref{tab:table_compare}.  For the SOLPS-ITER calculation for compression in Table \ref{tab:table_compare}, simulation values for helium density were taken at outer midplane, just outside separatrix since it is not possible for the SOLPS-ITER code to accurately calculate values within the domain of the core plasma. In the experiment, the values of the helium density used in the calculation for compression were measured at $\rho = 0.7$, roughly three times higher than value at separatrix. To compare the values of compression between experiment and simulation more precisely, the SOLPS-ITER value for helium density near the separatrix-core boundary used in the compression calculation was scaled by three to emulate the factor of three difference in the $\rho = 0.7$ and separatrix helium measurements. This scaled value is listed in the ``Corrected" row of Table \ref{tab:table_compare}. These modified compression values (listed in the ``Corrected" row of Table \ref{tab:table_compare}) are close in magnitude, indicating good agreement between experiment and simulation.

\begin{table}[]
\centering
\caption{Comparison of key helium exhaust parameters between DIII-D experimental measurements \cite{Wade_1998} and SOLPS-ITER simulations. The table lists the helium exhaust fraction, core helium fraction, helium enrichment factor, and helium compression. Values in the ``corrected" row represent compression recalculated or scaled to account for differences in measurement locations and domain limitations as described in the text, facilitating a more direct comparison. The tritium burn efficiency (TBE) is also included for reference.}
\begin{tabular}{|c|c|c|}
\hline
                      & DIII-D \cite{Wade_1998}  & SOLPS-ITER   \\ \hline
$f_{He,exh}$          & 6.1\%      & 6.0\%        \\ \hline
$f_{He,core}$          & 5.4\%      & 5.8\%        \\ \hline
$\eta_{He}$          & 1.1        & 1.0          \\ \hline
$\mathrm{C}_{He,exh}$ & 6.1 & 11.1 \\ \hline
Corrected $\mathrm{C}_{He,exh}$ & 4.4 & 3.8 \\ \hline
TBE                   & 0.1        & 0.1          \\ \hline
\end{tabular}
\label{tab:table_compare}
\end{table}

\subsection{Plasma parameters}
\begin{figure}
    \centering
    \includegraphics[width=\linewidth]{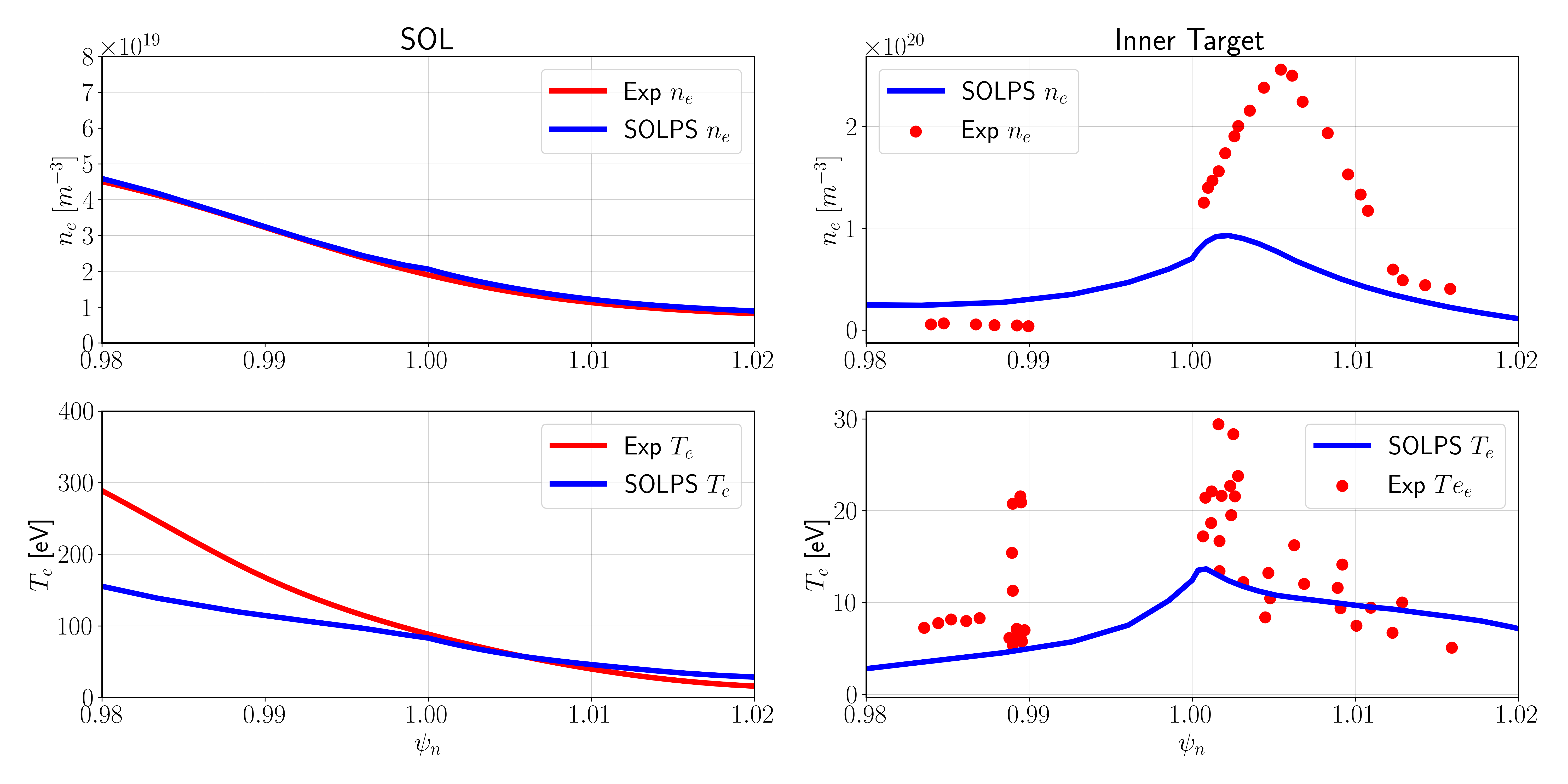}
    \caption{Midplane and target profiles in SOLPS (blue) and reproduced from publication \cite{Wade_1998} (red) for electron density (top row) and electron temperature (bottom row).}
    \label{fig:quad}
\end{figure}

Plasma parameters for the tuned simulation are shown in Figure \ref{fig:quad}.  Recreations of the data, extracted from the published figures, are shown as red marks on the upstream distributions for temperature and density, and the distributions for temperature and density at the inner target. With the tuning of the cross-field transport coefficients, shown in Figure \ref{fig:coefficients}, the upstream parameters achieved a good match for the density across the simulation domain, and the temperatures around and across the separatrix. The temperatures close to the core boundary of the domain are consistently too low compared to the available experimental data, but this was ignored since the fluid equations are not applicable in this region of the domain.

\begin{figure}
    \centering
    \includegraphics[width=0.5\linewidth]{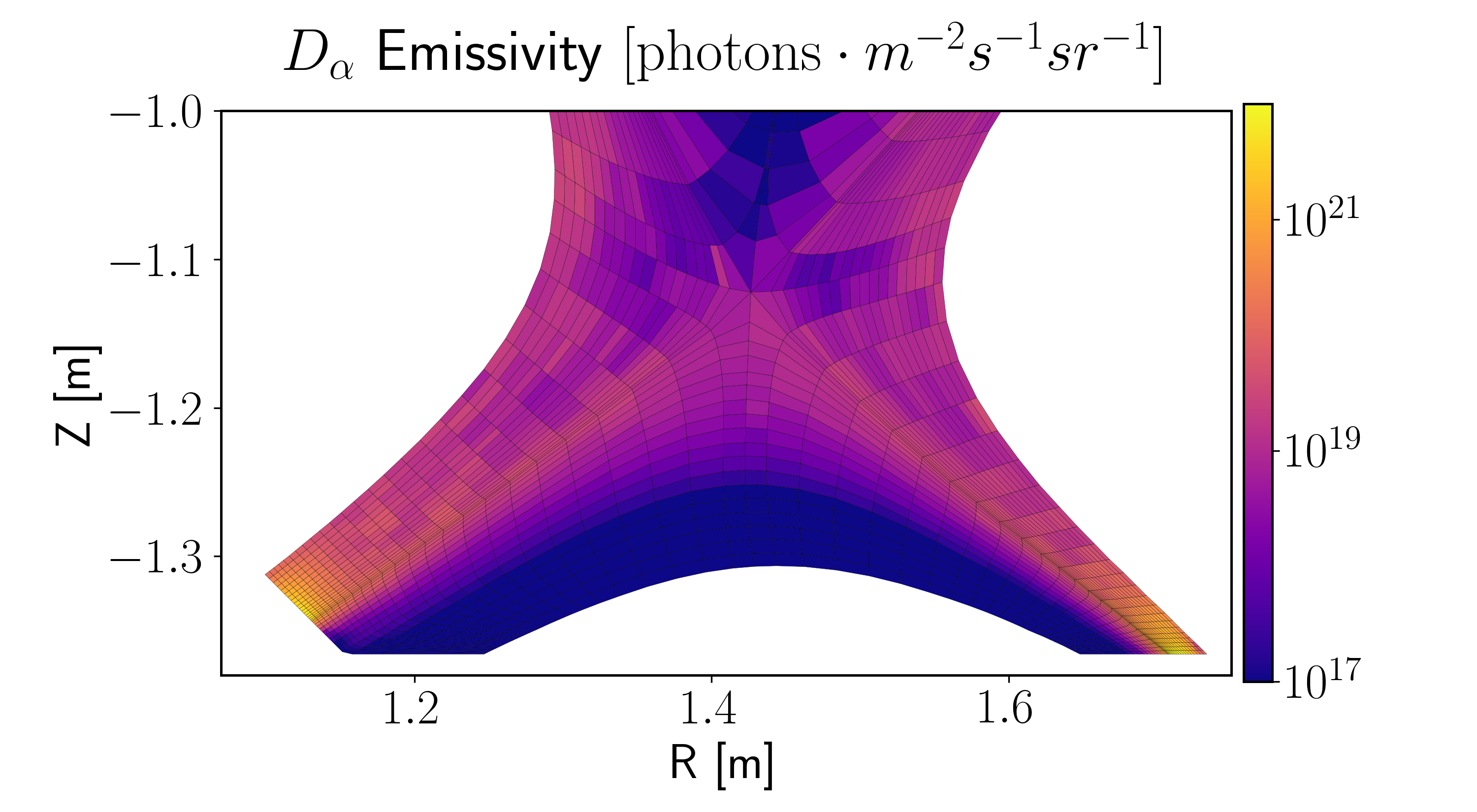}
    \caption{Comparison of simulated $D_\alpha$ emission with experimental measurements. The modeled $D_\alpha$ signal is on the same order of magnitude as the observed data, indicating reasonable agreement in the recycling dynamics. 
}
    \label{fig:d_alf}
\end{figure}

The upstream separatrix and inner target temperature and density at the inner target are fairly well-matched, with the simulation results displaying reasonable agreement with the published data. The simulated densities exhibit a modest systematic deviation toward lower values compared to the experimentally reported data. However, this discrepancy (likely) falls within the experimental uncertainty bounds of the probe and Thomson Scattering measurements \cite{chen2003langmuir,carlstrom1992}, though these error margins are neither depicted in the figure nor explicitly addressed in the original manuscript. The temperature at the inner target is a better match, coinciding with the data points available in the published data. While data is not available for the outer divertor target, the simulation results show values within a reasonable range compared to other simulations of DIII-D with similar experimental specifications. Additionally, the $D_{\alpha}$ measurement reported in the paper is of the same magnitude as that observed in the simulation (shown in Figure \ref{fig:d_alf}), which is consistent with the observation that both divertors are operating under attached, high-recycling conditions. Elevated $D_\alpha$ levels in both inner and outer divertor legs close to the target are consistent with an attached plasma in the high-recycling regime.

\subsubsection{2D Temperature and plasma density distributions}
\begin{figure}
    \centering
    \includegraphics[width=\linewidth]{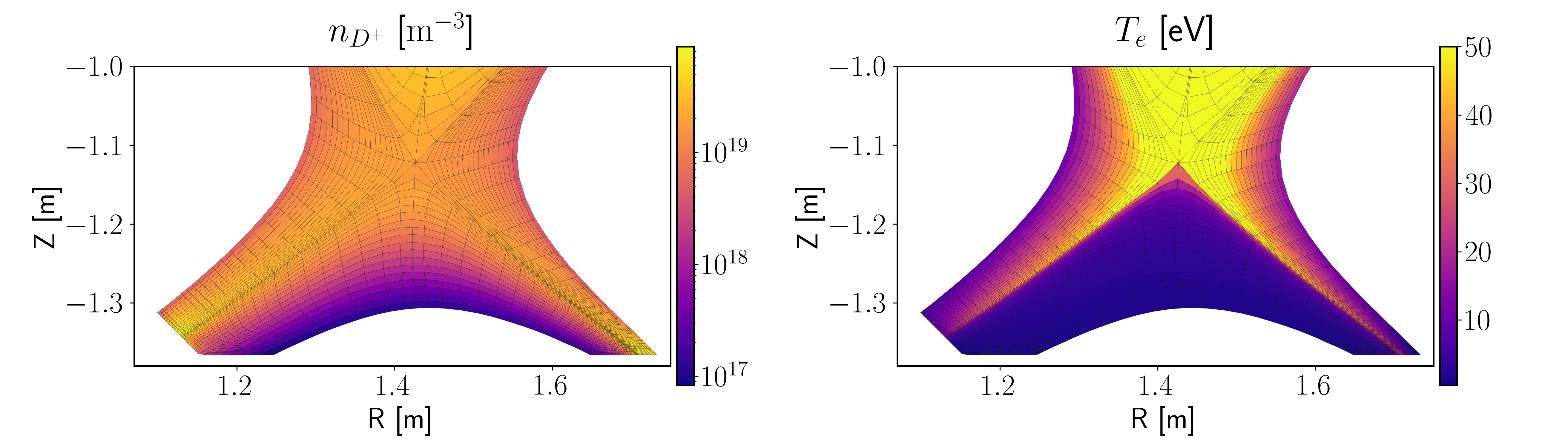}
    \caption{Two-dimensional profiles of deuterium ion density ($D^+$, left) and electron temperature ($T_e$, right). The $D^+$ density highlights the accumulation of plasma near the divertor targets, while the electron temperature profile shows elevated values near the separatrix and reduced (but still attached) temperatures toward the divertor plates.
}
    \label{fig:2dplasma}
\end{figure}

The temperature and density profiles in the divertor (shown in Figure \ref{fig:2dplasma}) strongly influence helium ionization states and transport dynamics, and therefore the overall distribution of helium in the divertor. Regions of low temperature but high density favor recombination and enhanced enrichment, while hotter regions near the separatrix tend to support higher ionization and faster parallel transport. These gradients determine where helium is likely to accumulate and whether it can be efficiently removed. 

\subsection{Neutral parameters}
\begin{figure}
    \centering
    \includegraphics[width=0.5\linewidth]{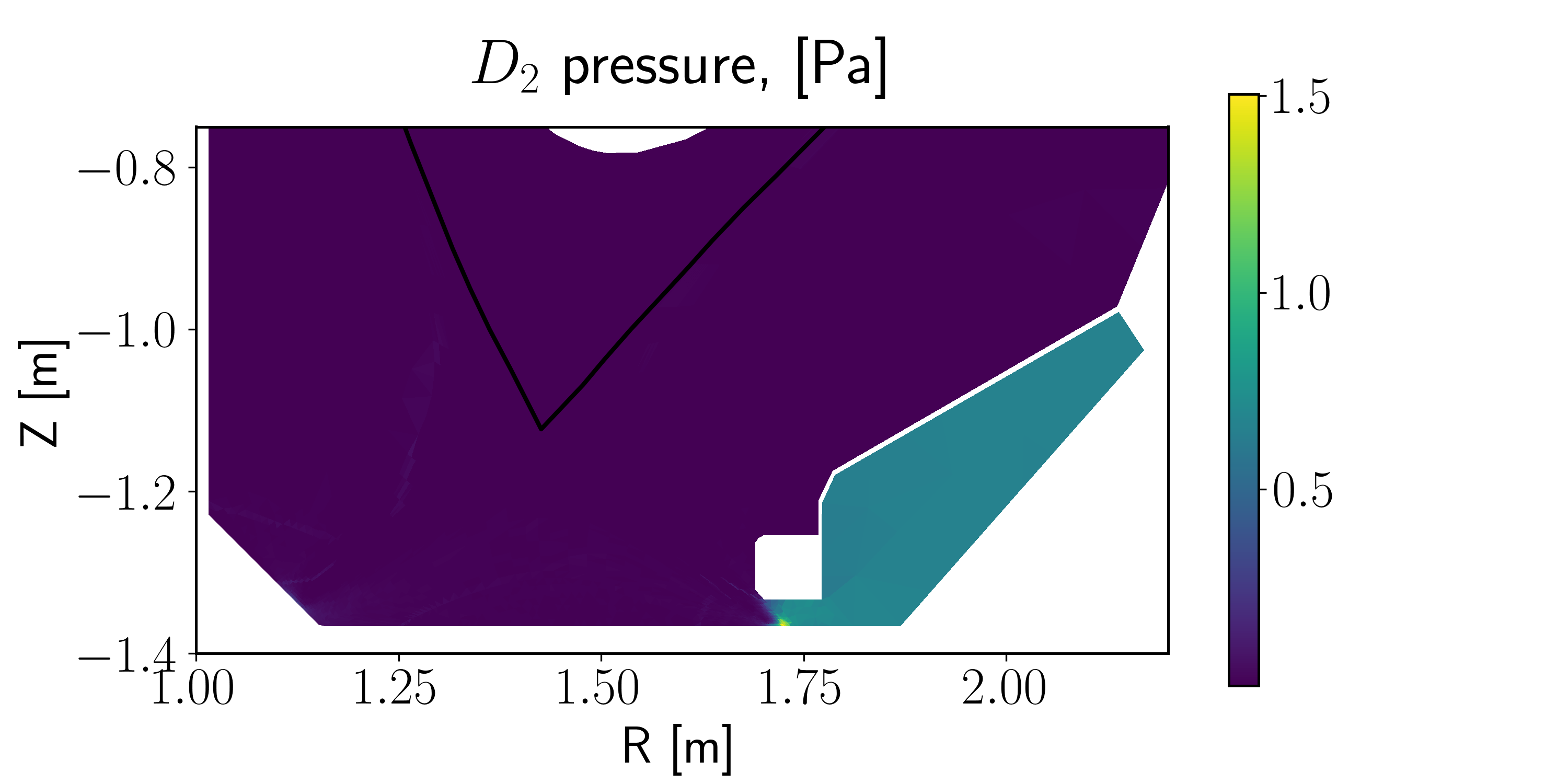}
    \caption{Distribution of molecular deuterium pressure in the divertor and plenum regions. The dark teal area represents the plenum, where the simulated pressure shows good agreement with experimentally measured value at 0.49 Pa, indicating a reasonable match between the modeled neutral dynamics and the diagnostic observations in this region.}
    \label{fig:d2pressure}
\end{figure}

Since special care was taken in the experiment to balance the SOL flows and removal of deuterium molecules in the pumping plenum, one of the key parameters matched in the simulation was the neutral pressure. This was achieved by fixing the pump absorption coefficient for $D_2$ removal to 2\% and 0\% for helium (meaning no helium is pumped out), and then adjusting the cross-field transport coefficient in the SOL and PFR components of the divertor volume (while leaving the tuned upstream radial diffusivity coefficients fixed) to impact the recycling dynamics in the divertor volume until the desired deuterium neutral pressure was achieved from the internal calculations in the EIRENE code. A match of 0.49 Pa, as reported in the paper, was achieved, shown in Figure \ref{fig:d2pressure}. The coefficients for helium were modified until the partial pressure of the helium in the exhaust plenum was approximately 6.1\% of the exhaust pressure, reported as $f_{exh}$ in the paper \cite{Wade_1998,west1999}. 

\section{Simulation Results and Analysis}
\label{results}

\subsection{Helium transport}
While this analysis does not provide a predictive model for helium transport, it does offer valuable insights through retrospective inference. By examining the adjustments to the radial transport coefficients required to achieve agreement between the simulation and the available experimental data, we can draw informed conclusions about the underlying transport behavior of helium in the divertor and near-separatrix regions. Although direct measurements of ion densities (such as those obtained via charge exchange recombination spectroscopy (CER) in the core plasma) are not available outside the separatrix, the consistency of the modeled density profiles with measurements inside the separatrix supports the validity of the inferred trends beyond it. In this context, we must rely on this physically grounded “educated guess” to extend our understanding into regions lacking direct diagnostic coverage, similarly to how transport profiles were determined from the MIST code in the initial study \cite{Wade_1998}.

One of the key features emerging from this matching process is the necessity to reduce diffusion transport coefficient for the helium species in the divertor and private flux regions compared to the radially-varying upstream values. This reduction was necessary to emulate the desired recycling dynamics in the divertor plasma volume that resulted in a good match to the reported helium pressure in the divertor plenum. The fact that a good match to the experimental profiles could only be achieved with suppressed transport suggests that helium is subject to enhanced confinement in the divertor relative to the upstream scrape-off layer, such that the parallel plasma flow patterns and corresponding local ionization balance are the dominant players in determining the retention mechanisms for the transport of the helium species, rather than the cross-field transport dynamics \cite{stangeby}. 

Ultimately, while the model does not predict helium behavior from first principles, the tuning process required to achieve alignment with experimental observations provides indirect but meaningful constraints on the transport mechanisms at play. The variation between the upstream and downstream characterization of the radial transport emphasizes the importance of localized transport physics in governing impurity behavior and can inform future model development or diagnostic strategies aimed at validating helium confinement predictions in reactor-scale devices, which will be the subject of future work.

\subsection{2D distributions}
\subsubsection{Helium distribution in divertor plasma}
\begin{figure}
    \centering
    \includegraphics[width=0.5\linewidth]{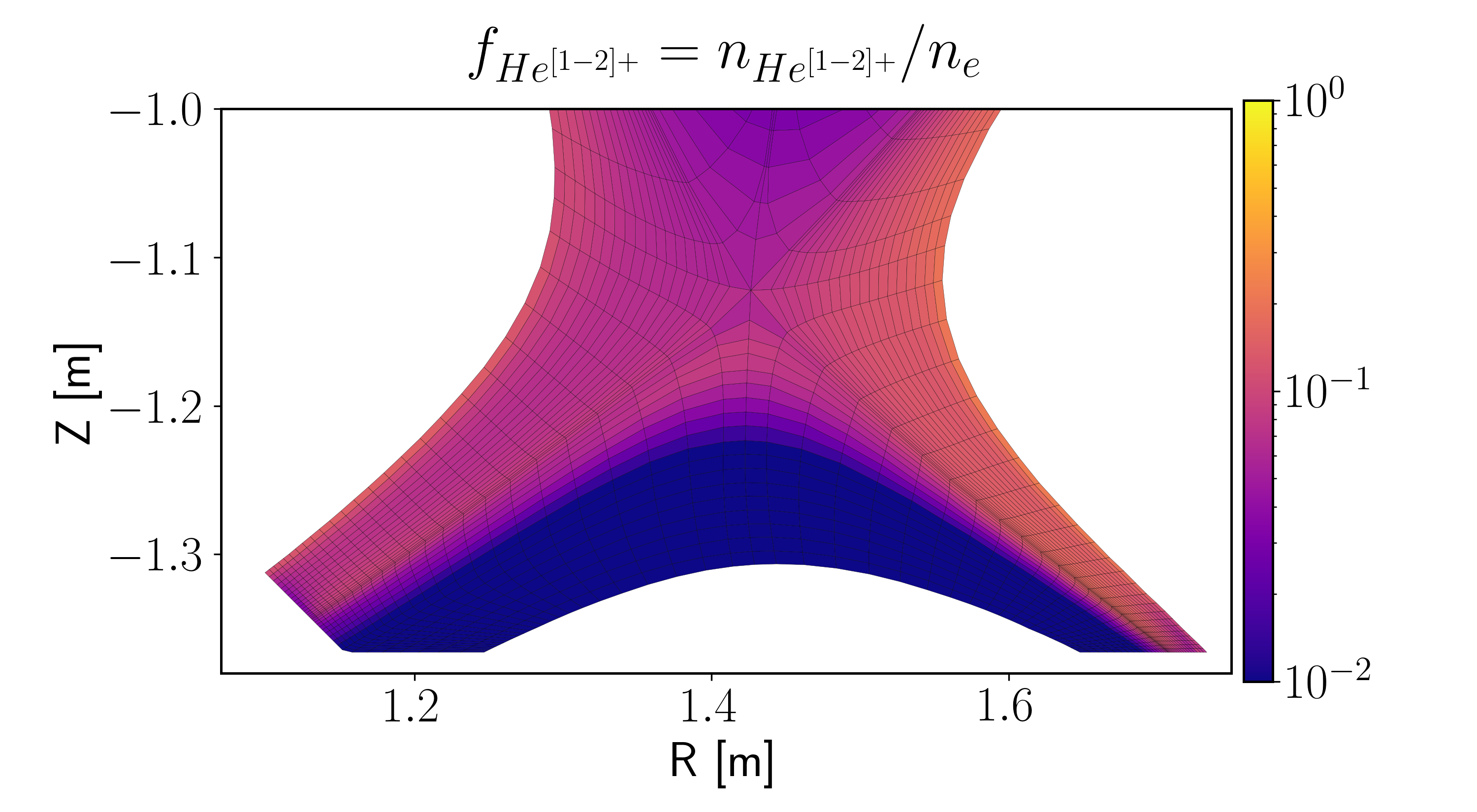}
    \caption{Relative density of ionized helium to electron density in each computational cell. This ratio provides a localized measure of impurity concentration throughout the domain, highlighting regions of helium accumulation relative to the bulk plasma. Values are elevated in the divertor region, consistent with effective helium confinement and recycling.}
    \label{fig:hefrac}
\end{figure}

\begin{figure}
    \centering
    \includegraphics[width=0.5\linewidth]{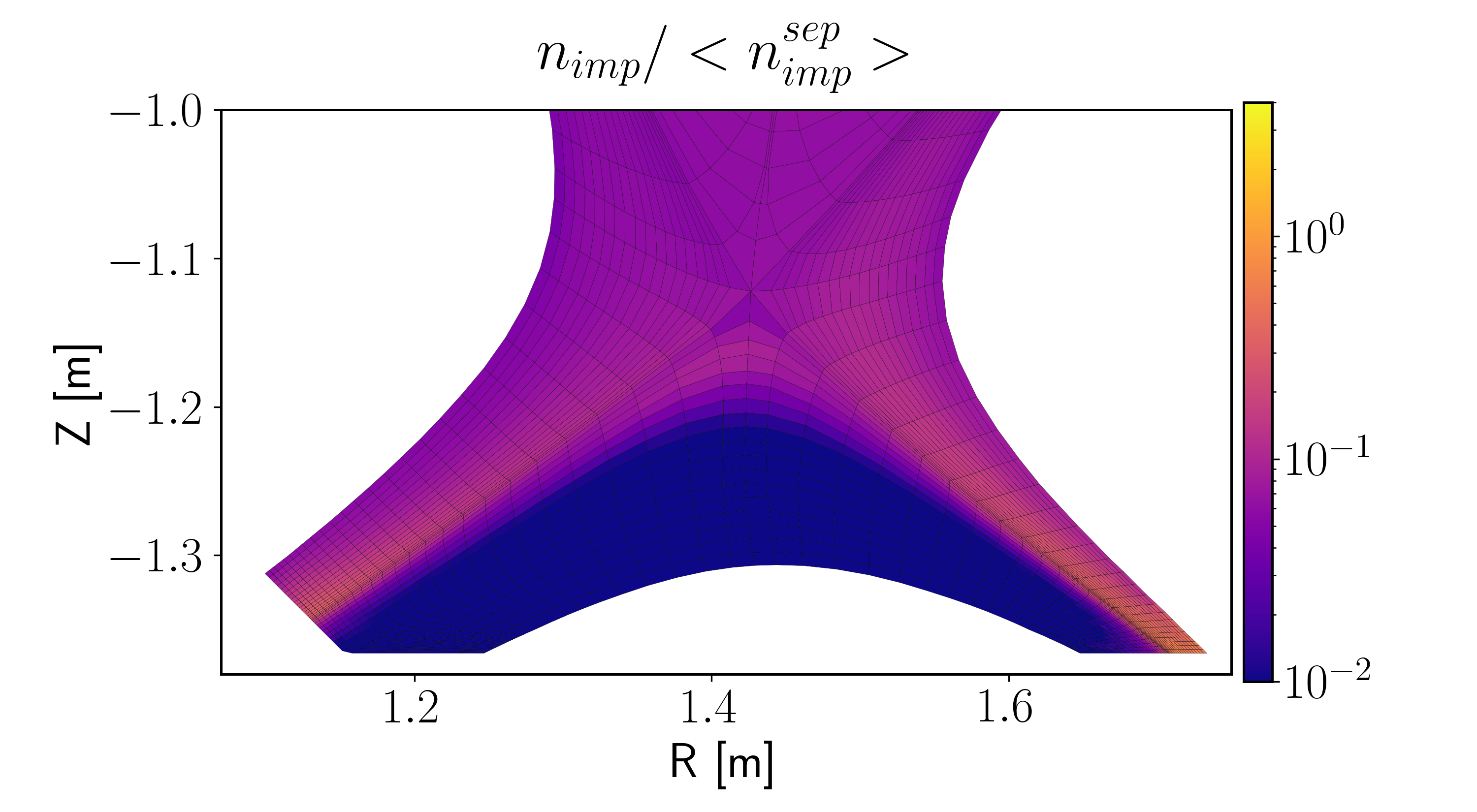}
    \caption{Volume-averaged compression of helium impurities, defined as the ratio of the total helium density in the divertor region to the helium density just above the X-point, serving as an indicator of impurity retention and divertor enrichment. Elevated compression values (seen along the separatrix in regions close to the targets) reflect effective confinement of helium within the divertor volume.}
    \label{fig:compression}
\end{figure}

\begin{figure}
    \centering
    \includegraphics[width=\linewidth]{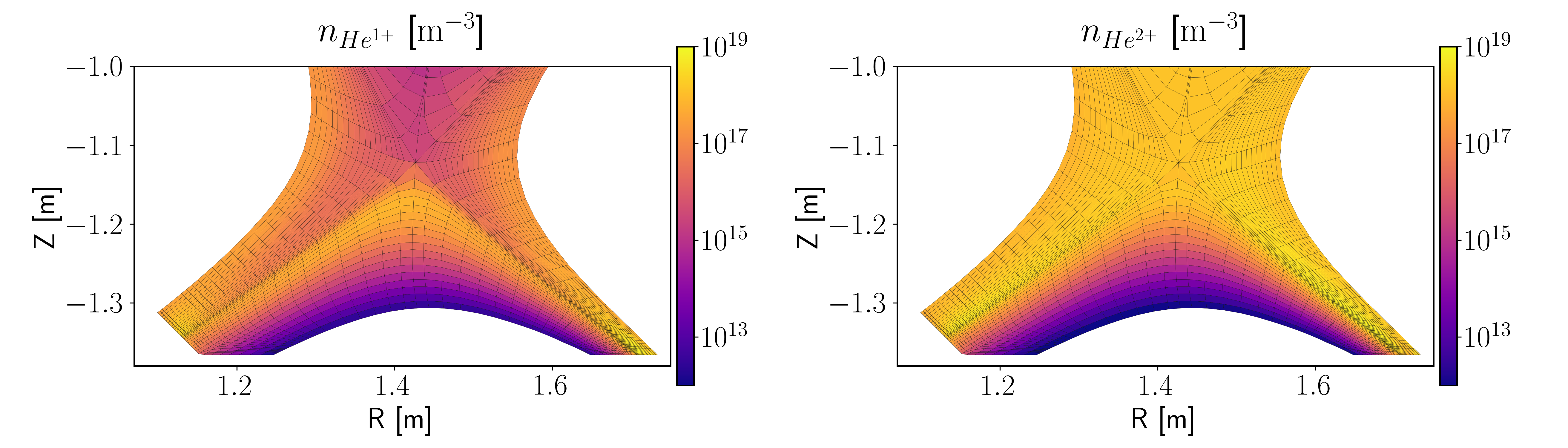}
    \caption{Two dimensional density distributions of helium in the divertor plasma. Left: Singly ionized helium ($He^{1+}$) is primarily concentrated near the divertor targets, reflecting prompt re-ionization of recycled neutral helium. Right: Fully ionized helium ($He^{2+}$) is localized closer to the separatrix, corresponding to regions of higher electron temperature where complete ionization is more likely to occur.}
    \label{fig:density}
\end{figure}

The helium distribution of both charge states of helium in the divertor plasma is shown in Figure \ref{fig:density}. Both charge states of helium, relative to the concentration of deuterium in each cell, is shown in Figure \ref{fig:hefrac}, and the compression is shown in Figure \ref{fig:compression}. 

The helium density, shown in Figure \ref{fig:density}, and relative density of helium in both ionization states ($\mathrm{He}^{1+}$ and $\mathrm{He}^{2+}$) compared to the local deuterium concentration, shown in Figure \ref{fig:hefrac}, provide insight into helium compression across the divertor and SOL. In regions near the divertor targets—especially on the outer leg—helium is well compressed, with total helium ion density reaching levels comparable to or exceeding those of deuterium. This reflects strong recycling and confinement of helium in the divertor volume. In the inner divertor, the helium-to-deuterium ratio remains elevated, though slightly lower than on the outer leg, consistent with lower plasma flow and neutral pumping efficiency. Moving upstream into the SOL and near the X-point, helium densities drop off more rapidly relative to deuterium, indicating limited cross-field transport and minimal leakage toward the core. The dominance of $\mathrm{He}^{2+}$ in most of the divertor volume also suggests that helium ions remain confined long enough to reach full ionization, further supporting the conclusion that helium is effectively compressed and retained in the divertor.

The compression of the impurity species in each computational cell is shown in Figure \ref{fig:compression}. Here, the methodology used to define impurity compression in this study is a novel adaptation of the approach introduced by \cite{SENICHENKOV2023} and used in \cite{sean_sparc}, defined as:

\begin{equation}
    \frac{n_{rad,imp}^{(div)}}{\langle n_{rad,imp}\rangle ^{(sep)}}
\end{equation}

where $\langle n_{rad,imp} \rangle^{(sep)}$ is the volume-averaged impurity radiation source density in the first continuous poloidal ring above the X-point, and $n_{rad,imp}^{(div)}$ is the cumulative density of all helium charge states in each cell. In both the inner and outer divertor legs, the fully ionized helium ($He^{2+}$) is found to be concentrated near the separatrix, regions of higher electron temperature. This localization reflects the sensitivity of helium ionization to thermal gradients, where higher temperatures promote the full ionization of helium. In contrast, the distribution of neutral helium tends to be broader and closer to the material surface, consistent with its origin from recycling processes and lower ionization potential.

A high compression value indicates that helium is being retained efficiently in the divertor volume, rather than leaking upstream into the scrape-off layer or core plasma. In these simulations, helium compression values are consistently elevated—often exceeding unity, highlighting the effectiveness of recycling, local ionization, and frictional coupling in concentrating helium in the divertor and increasing helium enrichment. 

\subsection{Helium recycling pattern}
\begin{figure}
    \centering
    \includegraphics[width=\linewidth]{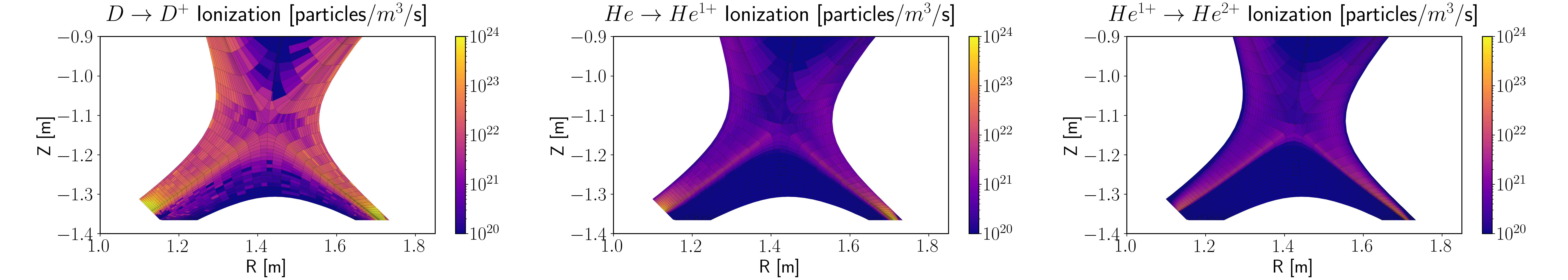}
    \caption{Two dimensional distribution of volumetric particle sources from ionization of recycling neutrals. Left: Deuterium ionization sources are broadly distributed throughout the divertor volume, consistent with volume ionization of recycled D neutrals. Center and Right: Helium ionization sources are sharply peaked near the divertor targets, indicating prompt re-ionization of recycled helium close to the wall due to its higher ionization energy and lower cross-section for charge exchange.}
    \label{fig:recycling}
\end{figure}

Helium recycling behavior, including the interaction of helium ions and neutrals with divertor targets and walls, plays a significant role in determining steady-state helium content. The pattern of recycling in the divertor, including the spatial localization of recycling sources and their impact on helium enrichment and transport, is determined by divertor geometry, target orientation, and energy balance in the divertor volume. The first ionization of the recycled helium particles (and deuterium particles) is shown in Figure \ref{fig:recycling}, showing strong ionization near the targets and in regions corresponding to high electron temperatures. However, unlike the deuterium ionization profile, which is peaked near the target but otherwise high throughout the divertor volume, the helium ionization profile is peaked at the target, with some ionization along the separatrix but otherwise much lower throughout the divertor volume. This highlights the difference in sources for $D^+$ and $He^{1+}$: deuterium ions primarily originate from ionization in the volume throughout the SOL, whereas singly ionized helium ($He^{1+}$) is mainly produced through ionization of recycled neutrals near the divertor target. 

This occurs because helium neutrals have a much higher ionization potential than deuterium, requiring hotter electrons to ionize. As a result, helium neutrals tend to penetrate deeper into the divertor and only ionize in regions of sufficiently high electron temperature, typically near the target surfaces. Deuterium, by contrast, has a much lower ionization threshold and can be ionized efficiently throughout the cooler regions of the divertor and SOL, resulting in a more spatially distributed ionization profile.

\subsubsection{Helium flux amplification}
Under the high recycling conditions in this discharge, there is considerable flux amplification of both the main ion species and the helium impurity, characterized in Table \ref{tab:fluxes}. The flux amplification ratio will be defined here as the ratio of the total particle flux of each species hitting the target plates relative to the total particle flux of each species entering the divertor leg. The fluxes entering the divertor leg (rather than the fluxes crossing the separatrix) are used to determine flux amplification in the divertor to eliminate main chamber recycling effects \cite{masline2023c} and the ionization of injected deuterium in the upper main chamber, shown in Figure \ref{fig:diiid-grid}. This captures both the plasma fluxes from the core and the fluxes from the ionization of deuterium molecules from the gas puff in the upper device.  

This improves divertor enrichment because the helium particles are undergoing multiple ionization and re-ionization cycles in the divertor volume without escaping to the main chamber. As these helium atoms and ions are trapped in recycling loops within the divertor, they accumulate and increase in density relative to their upstream concentrations. The repeated interactions with the dense, cold plasma and neutral background enhance the probability that helium is diverted toward the pumping region before it can return to the core. This process increases the local helium concentration in the divertor relative to upstream, raising the enrichment, and contributes to more effective helium exhaust by increasing the likelihood that helium will reach the plenum and be pumped out. Strong flux amplification, a hallmark of high-recycling regimes, supports preferential exhaust of helium through enhancing divertor retention of the helium species.

\begin{table}[]
\centering
\caption{Summary of particle fluxes across key boundaries in the SOLPS-ITER B2.5 computational mesh. Listed fluxes include those entering from core plasma, crossing the separatrix, impacting divertor targets, side boundaries, and entering or exiting through pumping surfaces. Fluxes are integrated over each boundary segment and represent the net transport of plasma and neutral species through the simulation domain. Positive values indicate net flux into the domain; negative values indicate net outflow.}
\begin{tabular}{|l|c|c|c|}
\hline
                             & \textbf{$D^{+}$} & \textbf{$He^{1+}$} & \textbf{$He^{2+}$} \\ \hline
\textbf{Flux across separatrix}       & 4.519e+22 & -2.237e+19 & 3.352e+20 \\ \hline
\textbf{Flux to main chamber wall}        & 8.292e+22 & 6.211e+18  & 2.332e+20 \\ \hline
\textbf{Flux to inner leg wall}    & 9.060e+20 & 1.246e+18  & 3.956e+18 \\ \hline
\textbf{Flux to outer leg wall}   & 2.109e+21 & 2.515e+18  & 1.302e+19 \\ \hline
\textbf{Flux through inner throat}      & 2.991e+21 & 1.594e+19  & 1.987e+20 \\ \hline
\textbf{Flux through outer throat}     & 2.591e+21 & 5.934e+17  & 1.646e+20 \\ \hline
\textbf{Flux to inner target}      & 2.304e+22 & 5.155e+20  & 4.259e+20 \\ \hline
\textbf{Flux to outer target}     & 3.349e+22 & 1.647e+21  & 6.053e+20 \\ \hline
\textbf{Total Particles In}        & 4.849e+18 & 6.377e+16  & 3.150e+17 \\ \hline
\textbf{Total Particles Out}       & 4.780e+18 & 9.224e+16  & 5.069e+17 \\ \hline
\textbf{Target/Core Ratio}   & 1.25      &            & 10.2      \\ \hline
\textbf{Target/Throat Ratio} & 10.13     &            & 8.41      \\ \hline
\end{tabular}%
\label{tab:fluxes}
\end{table}

\subsubsection{$\nabla \cdot \Gamma$ - Sources and sinks in flux tube}
\begin{figure}
    \centering
    \includegraphics[height=5cm]{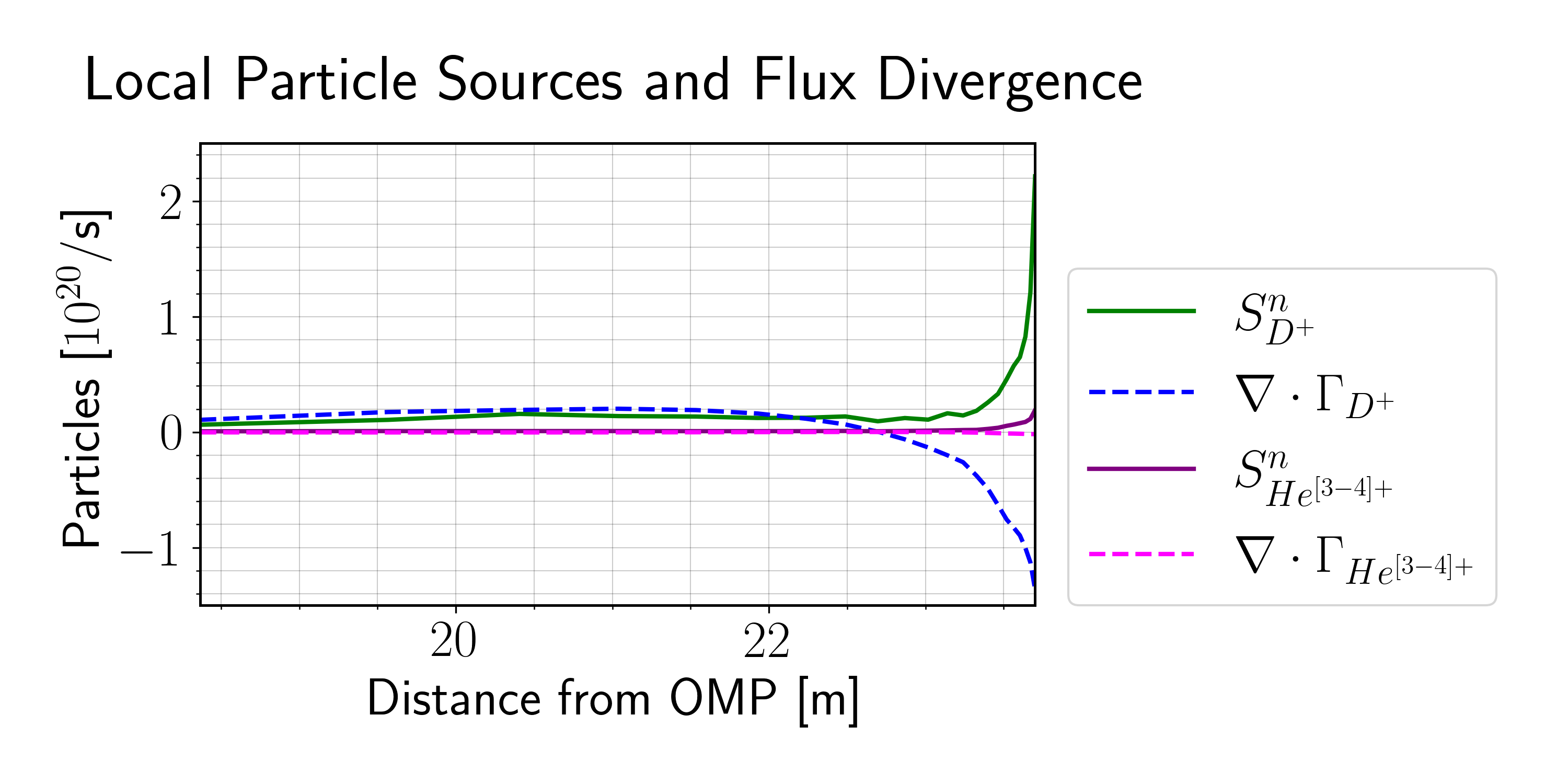}
    \caption{Spatial profiles of particle source terms (solid lines) and flux divergence (dashed lines) for both deuterium and helium species for the third flux tube from the separatrix the SOL. Green and blue curves correspond to the deuterium source and divergence, respectively, while purple and pink represent the source and divergence for singly and doubly ionized helium. This comparison highlights the balance between local sources (ionization and recycling) and sinks (cross-field transport) for each species, as well as differences in transport behavior and recycling dynamics between deuterium and helium.}
    \label{fig:paralleldiv}
\end{figure}

The net divergence of helium particle flux within a flux tube provides insight into the local balance of sources (ionization and recycling) and sinks (pumping and cross-field transport). Analyzing $\nabla \cdot \Gamma$ helps reveal where helium is produced, where it is lost, and how it redistributes along field lines in response to local plasma conditions. Figure \ref{fig:paralleldiv} shows the spatial distribution of the parallel flux divergence for a single flux tube, $\nabla \cdot \Gamma_\parallel$, for both the main ion species and the helium impurity. Positive values indicate net sources (regions where more particles are created than leave), while negative values indicate net sinks. As seen in the figure, the main ion species typically exhibit broader source regions due to widespread ionization of recycled deuterium throughout the divertor volume. In contrast, the helium species show more localized source regions near the divertor targets, reflecting their recycling and reionization behavior. These differences in source localization play a critical role in determining helium enrichment and retention in the divertor region.

It is notable that despite the strong ionization of the main plasma species, no flow reversal is observed in the divertor volume. Reversed parallel flow can arise in regions of strong ionization of the main plasma species, particularly when the local ionization source within a flux tube exceeds the ion flux reaching the divertor target, and results in a net flow of particles away from the target along the magnetic field line. This is a commonly observed feature in divertor plasmas \cite{Casali2022, Zito_2025}, which has a negative effect on impurity enrichment because it disrupts the directed transport of impurities toward the divertor targets and pumps. Impurity species, such as helium, tend to follow the background plasma flow due to frictional coupling with the main ions. When the main ion flow reverses direction within a flux tube, particularly upstream from the target in the divertor, impurities can be carried away from the divertor region and back toward the core or upstream SOL, reducing their residence time in the divertor volume (and limiting the recycling/reionization cycles that would impact pumping and removal) and increasing the likelihood that they escape into the main plasma.

However, such reversal does not necessarily occur in all high-ionization regions. In this case, although the divertor is in high recycling conditions with strong ionization for the main ion species, the strong induced parallel flow from the SOL means the background plasma flow dominates over localized source-driven dynamics. In these cases, even substantial ionization within a flux tube may not be sufficient to overcome the inertia of the background flow, and the overall particle transport remains directed toward the divertor targets. This illustrates the importance of global plasma dynamics in determining local flow structures, since strong main ion flow can suppress or mask localized flow reversals that might otherwise develop in response to steep ionization gradients alone. This strong flow keeps the recycling of the helium species more localized near the targets, which is favorable for enrichment.

\subsection{Stagnation point}
A flow stagnation point, where parallel plasma flows slow or reverse, is a known feature in some experimental divertor conditions and can contribute to impurity accumulation. The presence or absence of such a feature in experiments and simulations has a strong influence on helium enrichment. The ``puff and pump" experiment, with strong induced SOL flows, provides unique conditions that largely eliminate any flow reversal for the helium species and enable conditions that are favorable for helium enrichment in the divertor. In this section, the influence of the stagnation surface and resultant implications are discussed.

\subsubsection{Velocity field}
\begin{figure}
    \centering
    \includegraphics[width=\linewidth]{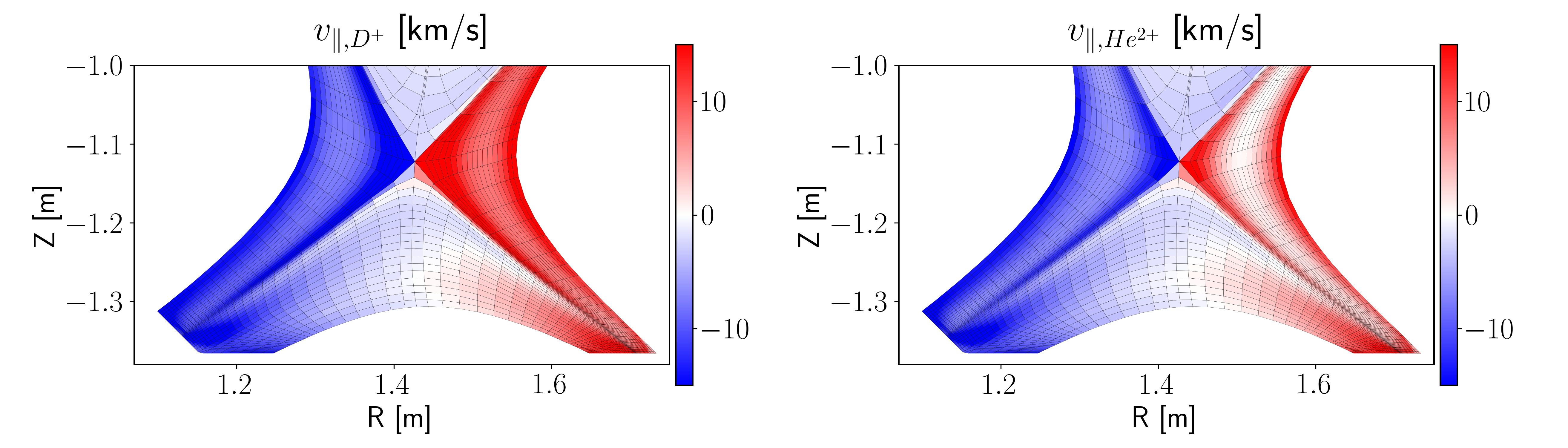}
    \caption{Two-dimensional parallel velocity fields on the simulation grid for the plasma main ion species (left) and helium species (right). In both cases, red indicates flow directed toward the outer divertor target (positive along the field line), and blue indicates flow toward the inner target (negative). No stagnation surface where the parallel flow direction reverses is observed in the divertor volume for either species, indicating continuous flow toward the respective targets throughout the flux tubes.}
    \label{fig:velocity}
\end{figure}

Mapping the full plasma and impurity parallel velocity fields provide a global view of the flow structure that governs impurity motion, since the presence of a ``stagnation surface" in the divertor volume with an associated flow reversal region where the direction of impurity flow reverses is a major determiner of the enrichment of the impurity in the divertor. Differences between helium and main ion velocities can indicate frictional forces or differential confinement that affect helium transport and removal from the divertor volume. The velocity field for both the plasma and fully ionized helium $He^{2+}$ impurity species is shown in Figure \ref{fig:velocity}. Regions shaded in a blue hue are indicative of velocities directed towards the inner target, while regions shaded in a red hue are directed towards the outer target, and white regions represent regions of flow stagnation. 

\subsubsection{Velocities in flux tube}
\begin{figure}
    \centering
    \includegraphics[height=5cm]{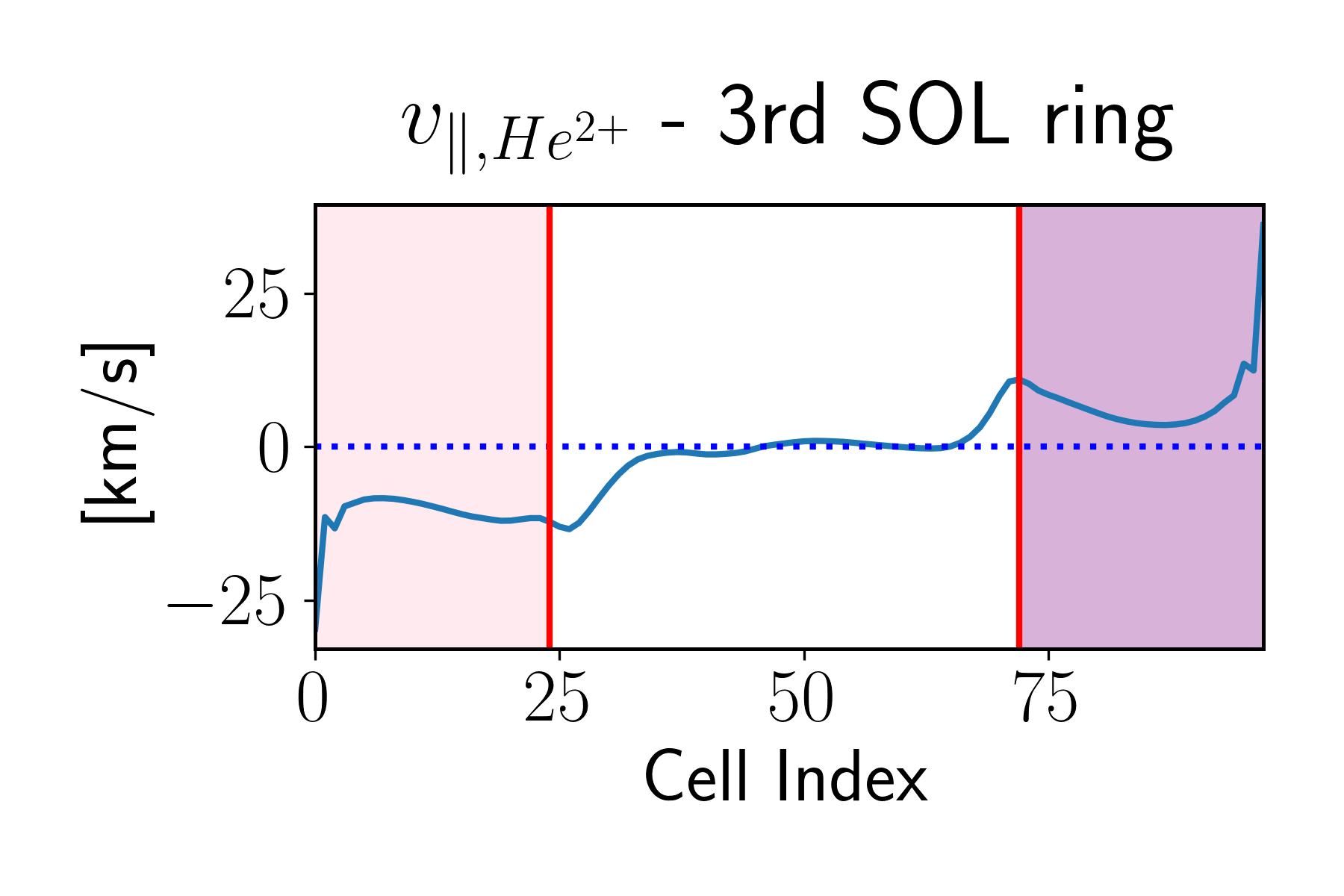}
    \caption{Parallel flow velocity in the divertor region, where pink and purple shaded regions represent the inner and outer divertor legs, respectively. The red region marks the divertor throat. A stagnation point would correspond to a location where the parallel velocity crosses zero, indicated by the dotted contour line. In this case, no such crossing is observed within the divertor volume, indicating that no stagnation point is present. Flow remains directed toward the respective divertor targets throughout the region.}
    \label{fig:velocity_ft}
\end{figure}

Velocities in a near-SOL flux tube (third ring from the separatrix) are shown in Figure \ref{fig:velocity_ft}, where negative values indicate velocities directed toward the inner target along the parallel magnetic field line, while positive values correspond to velocities toward the outer target. As seen in the figure, no stagnation point is observed in either divertor; the helium velocities remain consistently directed toward their respective targets. This unidirectional flow suggests that, in this region, there is no local accumulation or trapping of helium, and that transport is governed primarily by parallel friction and pressure gradients rather than any reversal or stagnation of flow.

\subsubsection{Force balance in flux tube}

In the fluid approximation, the simplified equation for the force balance on impurity ions is \cite{stangeby}

\begin{equation}
    F_z = - \frac{1}{n_z} \frac{dp_z}{ds} + m_z \frac{(v_i - v_z)}{\tau_s} + ZeE + \alpha_e \frac{d(kT_e)}{ds} + \beta_i \frac{d(kT_i)}{ds},
    \label{eq:force}
\end{equation}

where each of these terms in the the impurity momentum balance describes the four primary parallel forces acting on the impurity ions: the impurity pressure gradient force, the friction force from main ions, the electrostatic force from the parallel electric field, and the temperature gradient force (composed of an electron and an ion component). These forces collectively determine the direction and strength of impurity transport along magnetic field lines in the scrape-off layer (SOL).

\begin{figure}
    \centering
    \includegraphics[width=\linewidth]{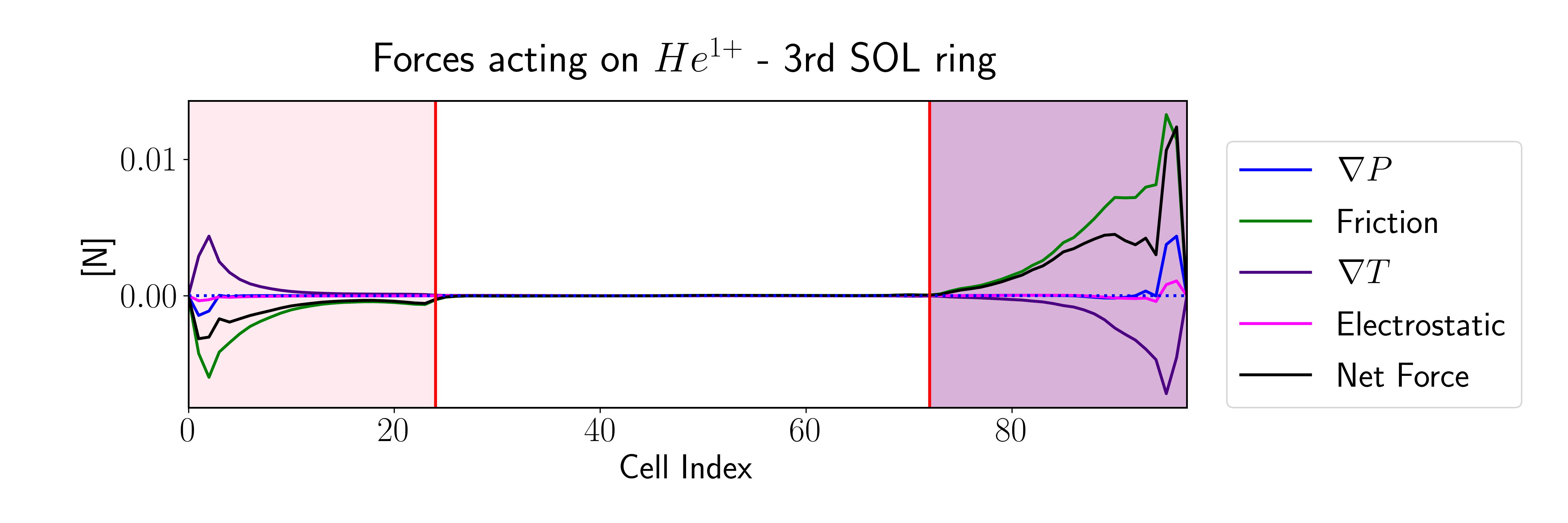}
    \caption{Force balance components acting on singly ionized helium (He$^{1+}$) within a single flux tube from the third SOL ring of the computational domain, where pink and purple shaded regions represent the inner and outer divertor legs, respectively. The pressure gradient force is shown in blue, the friction force in orange, the ion temperature gradient force in green, and the electrostatic force in red. This plot illustrates the relative contributions and spatial variation of each force term influencing helium transport along the field line in the divertor region, starting at the inner target (index 0) to the outer target (index 99).}
    \label{fig:forces}
\end{figure}

The force balance (with each of these four forces) acting on singly ionized helium ions in the third SOL ring from the separatrix shown in Figure \ref{fig:forces}. Here, negative values indicate forces directed toward the inner target along the parallel field line, while positive values correspond to forces pointing toward the outer target along the parallel field line. Across both inner and outer divertors, the dominant contributions to the force balance are the friction force (arising from collisions with the bulk plasma) and the ion temperature gradient force. These forces govern the direction and magnitude of helium transport in the SOL, with strong main ion flow playing a critical role in confining helium within the divertor region by advecting impurities toward the targets and suppressing upstream leakage toward the core.

\subsubsection{Induced flow increasing divertor enrichment and preventing leakage}
Strong induced parallel flows, especially in high-fueling scenarios like the one modeled in this work, enhance helium confinement in the divertor by increasing friction forces between the main species and impurity and reducing leakage into the core plasma. However, generating such flows typically requires increased deuterium fueling, which is in tension with the need to minimize tritium inventory. Strategies for maintaining high tritium burn efficiency (TBE) while sustaining effective helium exhaust, such as advanced pumping technology or alternative fueling schemes like spin polarization, are critical for optimizing reactor performance.

\section{Tritium Burn Efficiency}
\label{tbe}

The application of the TBE parameter to edge plasma simulations assumes a fixed level of fuel dilution in the upstream plasma. This assumption is not self-consistent within the SOLPS-ITER framework, but is reasonable in the context of existing models that include a prescribed degree of fuel dilution. In reality, the helium flux from the core would depend on the evolving dilution of the main fuel, which in turn would affect both the impurity source and the power balance. However, the TBE method does not capture this feedback. Instead, it should be interpreted as representing a self-consistent steady-state scenario: given a specific helium particle flux and associated power input, a plasma with these edge characteristics \textit{can} exist. This is distinct from a predictive approach where a given power level and dilution are assumed to deterministically yield a corresponding helium flux. Thus, for this study, the application of TBE to the simulation results is presented as characterizing a consistent snapshot of the edge plasma under specified conditions, rather than a dynamic outcome of core-edge coupling (which is beyond the scope of this work).

The goal of this study is to leverage existing experimental data in order to better characterize and understand the behavior of realistic divertor plasmas with helium to assess the viability of the TBE parameter - to do this, it is necessary to understand and interpret the quantities in the TBE metric and verify whether they are consistent between experiment and the available quantities in SOLPS-ITER. In other words, given the helium fraction in the divertor, the pumping speed ratio, and the resultant burn efficiency itself, we aim to assess whether it would be reasonable to assume that this edge plasma state (which will have realistic compression and enrichment values that can characterize dilution of the near-SOL simulation domain) could be consistent with a realistic core plasma.

\subsection{Helium fraction - $f_{He}$}
\begin{figure}
    \centering
    \includegraphics[width=0.5\linewidth]{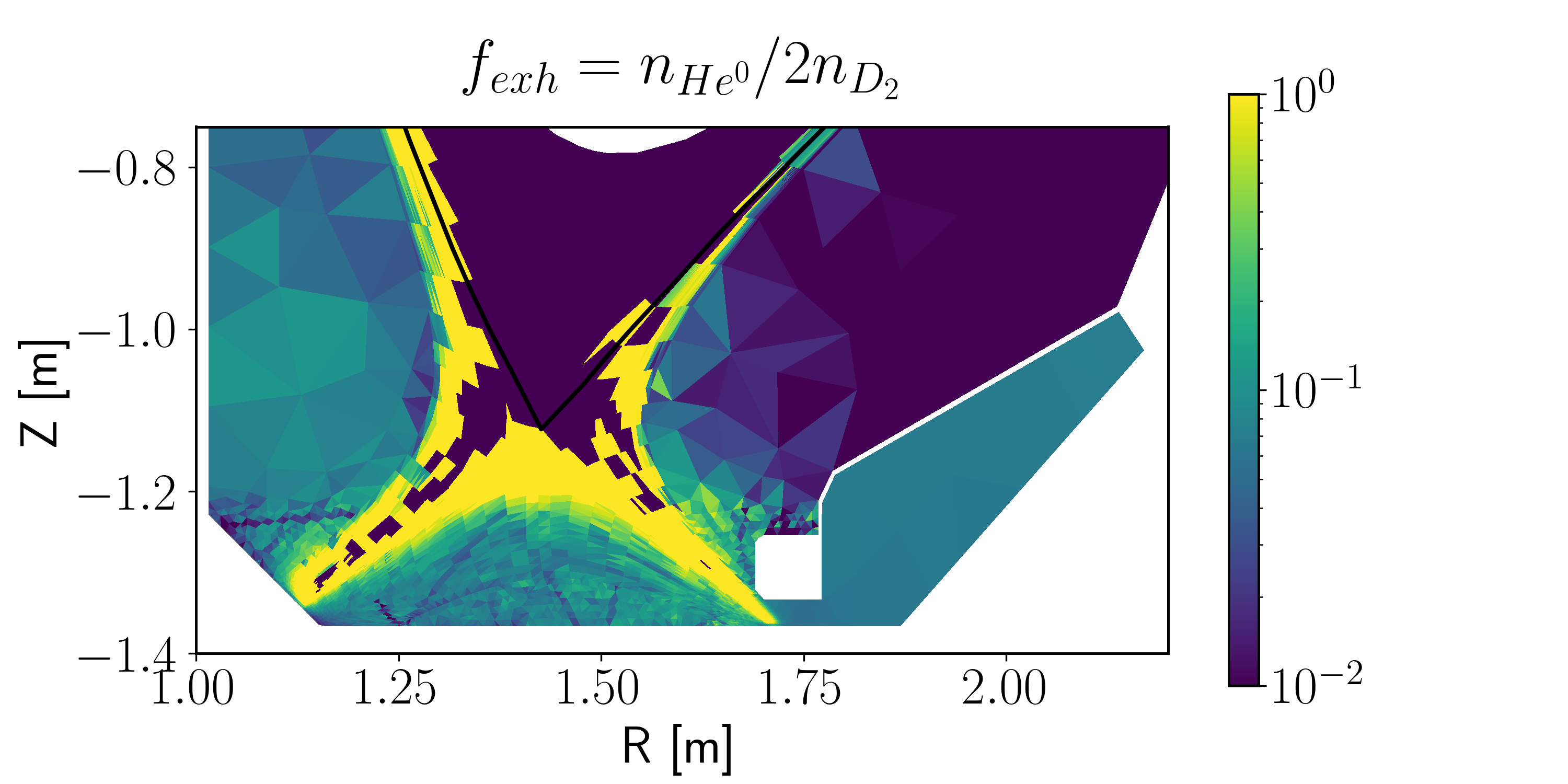}
    \caption{Exhaust fraction in the divertor region, defined as the ratio of neutral helium density to twice the molecular deuterium density. This metric, which is used in the calculation for TBE, provides insight into helium removal efficiency relative to the dominant neutral hydrogenic background. Yellow regions indicate areas with negligible or zero molecular deuterium density, where the ratio is undefined or not meaningful.
}
    \label{fig:fexh}
\end{figure}

The helium fraction, $f_{He}$, defined as the ratio of helium ion density to total ion density, serves as both a marker of successful fusion reactions in the TBE frame work and a constraint on plasma performance due to fuel dilution (which would, in turn, reduce TBE). Thus, the goal is to strike a balance: maintaining a low $f_{He}$ in regions where fuel purity is critical, like the core plasma, while recognizing that some helium presence is a sign of effective burning, since helium cannot be removed immediately as it is generated. Achieving this balance hinges on the ability to preferentially exhaust helium from the edge plasma without significantly affecting the hydrogenic species. In practice, this requires either highly effective pumping systems that are capable of selectively removing helium or divertor geometries and flow structures that naturally favor helium exhaust. Optimizing these factors is key to sustaining high-performance, reactor-relevant plasmas without sacrificing TBE. For this DIII-D simulation, a divertor plasma with helium fraction of around 6\% is achieved (shown in Figure \ref{fig:fexh}), with an enrichment value of around 1 and a core helium fraction of around 5\%. This is already within the range of a fuel dilution of 10\% considered acceptable for ITER \cite{Gilleland_1989}, but it is noted that there is considerable room for optimization with respect to helium exhaust in the DIII-D divertor plasma: the divertor is open, the targets are flat, and there was no actual helium exhaust in this experiment, all of which can improve the efficiency of helium exhaust.  

\subsection{Pumping speed ratio - $\Sigma$}
In the absence of dedicated experiments on helium enrichment, Section 3.3 of \cite{tritium_memo} offers context by estimating a plausible range of TBE using published experimental data. These historical experiments did not include direct helium pumping due to the technological limitations of helium removal using cryopumps, so a default value of $\Sigma = 1$ is assumed in the TBE calculations. This assumption implies that the effective pumping speed for helium is equal to that of the deuterium species. Physically, this means helium and deuterium would be removed from the system at the same rate, neglecting any real-world differences in species-dependent pump efficiency or pumping conductance that might preferentially retain helium. While simplifying, this assumption provides a useful baseline for comparing achievable TBE in systems without advanced helium exhaust capabilities, like this one.

\subsection{TBE}
With the assumptions made given the constraints of the available pumping technology for DIII-D, a TBE of approximately 0.1 is achievable in realistic divertor plasmas. Given the extreme scarcity of tritium and the stringent requirement for tritium self-sufficiency in a D–T fusion pilot plant, achieving a TBE of approximately 0.1 is a promising result. This means that at steady state, 10\% of the tritium injected into the plasma would be burned, significantly reducing the burden on the tritium breeding and recovery systems. Even modest improvements in TBE translate directly into lower requirements for breeding blanket coverage, shorter tritium processing timescales, and smaller on-site inventory needs, each of which contributes to the overall feasibility and safety of tritium fuel cycle management in an FPP. This level of efficiency underscores the need to improve helium removal to prevent dilution and enhance TBE through both technological and design advances such that the pumping efficiency of helium is equivalent to that of the bulk species to be able to achieve this in reality without making necessary assumptions. Cryopumps used in many current tokamaks are effective for pumping hydrogen isotopes, but tend to have poor performance for helium. Addressing this limitation may involve the development of alternative pump technologies tailored for helium capture, such as metal-based getter pumps or high-capacity turbo-molecular systems \cite{BAQUERORUIZ2021,Zito_2025}. In parallel, optimizing divertor geometry and pump placement to enhance removal of the helium species can significantly improve helium exhaust without adversely affecting core confinement. This emphasizes the importance of an integrated approach to impurity control in future fusion devices, where helium management will be central to sustaining high-efficiency operation and tritium self-sufficiency.

\section{Conclusions}
\label{conclusions}
Helium enrichment in the divertor appears to be reasonably well captured by SOLPS-ITER simulations under conditions where friction with the bulk plasma is the dominant enrichment mechanism. In this regime, parallel frictional forces arising from interactions between helium ions and the background deuterium plasma act to impede the escape of helium from the divertor region, effectively localizing it. This behavior is consistent with scenarios where other enrichment processes, such as neutral trapping or wall pumping, play a minimal role, since they were not included in the simulation.

To achieve quantitative agreement with helium density measurements, however, a significant reduction in the effective cross-field transport coefficient for helium diffusivity was required in the simulation. Physically, this implies that helium is less mobile across magnetic flux surfaces in the divertor than standard upstream transport assumptions would suggest. The necessity of this tuning itself also highlights the non-uniformity of impurity transport across different plasma regions and underscores the need for region-specific transport models.

In contrast, the work presented in \cite{Zito_2025} points to significant discrepancies between simulation and experiment in similar regimes, but without strong induced flows of the main plasma species and the presence of a clear stagnation surface located in the divertor volume. This suggests that ionization balance in existing edge codes may not be accurately resolved, and that without correctly capturing the ionization dynamics, edge codes seem to under-predict helium concentrations in the divertor. This difference is critical, as the stagnation point alters the parallel flow structure and enhances helium accumulation, an effect that is not captured in simulations that lack similar flow features to that of the DIII-D simulation.

Lastly, tritium burn efficiency (TBE) is a valuable metric for assessing tritium fuel utilization in reactor-scale devices, as it is closely linked to the helium fraction in the edge and divertor plasma. While a TBE of 0.1 appears achievable based on divertor plasma conditions, realizing this in practice would require advanced pumping technologies beyond the capabilities of most current tokamaks, underscoring the need for optimized pumping systems.

\section*{Acknowledgements}
This research was supported by the U.S. Department of Energy (DOE) Fusion Energy Sciences Postdoctoral Research Program administered by the Oak Ridge Institute for Science and Education (ORISE) for the DOE. ORISE is managed by Oak Ridge Associated Universities (ORAU) under DOE contract number DE-SC0014664. All opinions expressed in this paper are the author's and do not necessarily reflect the policies and views of DOE, ORAU, or ORISE.

The author gratefully thanks T. Eich, A. Creely, T. Body, P. Bonoli, P. Catto, and S. Frank for their support during the preparation of this work.

\section*{References}
\bibliographystyle{unsrt}
\bibliography{RefList.bib}

\end{document}